\documentclass[12pt,a4paper]{article}
\usepackage{graphicx}
\usepackage{times}
\title{\bf REMAP: Determination of the inner edge of the dust torus in AGN
by measuring time delays\footnote{Echo Mapping of Active Galactic Nuclei}}

\author{Amit Kumar Mandal $^{1,2}$\thanks{E-mail: amitkumar@iiap.res.in} , Suvendu Rakshit $^3$, Indrani Pal $^2$, 
Chelliah Subramonian Stalin $^2$, \\ Ram Sagar $^2$, 
Blesson Mathew $^1$\\
\vspace{0.5cm}\\
\normalsize $^1$ Department of Physics and Electronics, CHRIST (Deemed to be University),\\ Hosur Road, Bangalore 560 029, India \\
\normalsize $^2$ Indian Institute of Astrophysics, Block II, Koramangala, Bangalore 560 034, India \\
\normalsize $^3$ Finnish Centre for Astronomy with ESO (FINCA),\\ University of Turku, Quantum, Vesilinnantie 5, 20014, Finland \\ }
\date{\mbox{}}
\begin{document}
\maketitle
\setcounter{page}{1001}
\pagestyle{plain}
    \makeatletter
    \renewcommand*{\pagenumbering}[1]{%
       \gdef\thepage{\csname @#1\endcsname\c@page}%
    }
    \makeatother
\pagenumbering{arabic}

%
%
\def\bull{\vrule height .9ex width .8ex depth -.1ex}
\makeatletter
\def\ps@plain{\let\@mkboth\gobbletwo
\def\@oddhead{}\def\@oddfoot{\hfil\scriptsize\bull\quad
"2nd Belgo-Indian Network for Astronomy \& astrophysics (BINA) workshop'', held in Brussels (Belgium), 9-12 October 2018 \quad\bull}%
\def\@evenhead{}\let\@evenfoot\@oddfoot}
\makeatother
%
%
\def\beginrefer{\section*{References}%
\begin{quotation}\mbox{}\par}
\def\refer#1\par{{\setlength{\parindent}{-\leftmargin}\indent#1\par}}
\def\endrefer{\end{quotation}}
%
%
{\noindent\small{\bf Abstract:} 
Active galactic nuclei (AGN) are high luminosity sources powered by accretion 
of matter onto super-massive black holes (SMBHs) located at the centres of 
galaxies. According to the Unification model of AGN, the SMBH is surrounded by a broad emission line region (BLR) and a dusty torus. It is difficult to 
study the extent of the dusty torus as the central region of AGN is not 
resolvable using any conventional imaging techniques available today. Though, current 
IR interferometric techniques could in principle resolve the torus in 
nearby AGN, it is very expensive and limited to few bright and nearby AGN. A more feasible alternative to the interferometric technique to find 
the extent of the dusty torus in AGN is the technique of reverberation 
mapping (RM). REMAP (REverberation Mapping of AGN Program) is a long term 
photometric monitoring program being carried out using the 2 m Himalayan 
Chandra Telescope (HCT) operated by the Indian Institute 
of Astrophysics, Bangalore, aimed at measuring the torus size in many AGN using
the technique of RM. It involves accumulation of suitably long and 
well sampled light curves in the optical and near-infrared bands to measure
the time delays between the light curves in different wavebands. These 
delays are used to determine the radius of the inner
edge of the dust torus. REMAP was initiated in the year 2016 and since then about 
one hour of observing time once every five days (weather permitting) has been 
allocated at the HCT. Our initial sample carefully selected for this program 
consists of a total of 8 sources observable 
using the HCT. REMAP has resulted in the determination of the  
extent of the inner edge of the dusty torus in one AGN 
namely H0507+164. Data accumulation for the second source is completed and
observations on the third source are going on. We  will outline the motivation 
of this observational  program, the observational strategy that is followed, 
the  analysis procedures adopted for this work and the results obtained 
from this program till now.
}
\vspace{0.5cm}\\
{\noindent\small{\bf Keywords:} galaxies -- active -- quasars -- Seyfert -- reverberation}
%
%
\section{Introduction}
The unified model of active galactic nuclei (AGN; Urry $\&$ Padovani 1995; 
Antonucci 1993) 
posits a dusty torus surrounding the broad line region (BLR) and the central super 
massive black hole (SMBH). 
The broad band spectral energy distribution (SED) 
of AGN contains a prominent big blue bump indicative of emission from the 
accretion disk and an excess emission in the infrared (IR) region attributed to 
thermal emission from the dusty torus. The dusty torus absorbs the UV/optical 
radiation from the accretion disk and re-emits it in the IR band. The torus is 
therefore thought to be the dominant source of IR radiation in most AGN and can 
also explain the IR bump seen in the SED of AGN. Understanding this obscuration 
of the central engine is therefore important to constrain the physical 
processes happening in the central regions of an AGN.  However, the torus is so 
small that it is not possible to image it using        
any existing imaging system on a single telescope.  
In spite of our inability to directly image the torus using a 
single telescope, it is possible to infer the size and structure of the
torus via two important methods namely (a) reverberation mapping (RM; 
Blandford $\&$ McKee 1982) and
(b) IR interferometry. In the method of RM,  if we monitor 
AGN continuum fluxes in the UV/optical 
and near-infrared, a time-lag would be expected that corresponds to the 
distance between the central engine and the hot dust region from which the bulk 
of near-infrared emission is radiated (e.g., Clavel, Wamsteker \& Glass 1989; 
Nelson 1996; Glass 2004). By measuring this time delay one can (i) put 
constraints on the inner radius of the dusty torus that surrounds the AGN and 
(ii) probe the physical processes happening in the central regions of 
an AGN. Such reverberation lags between near-IR (K-band) and optical
V-band light curves are available for about 20 Seyfert galaxies and these lags are believed to represent the inner radius (R$_{in}$) of the torus
(Minezaki et al. 2004; 
Suganuma et al. 2006; Koshida et al. 2009, 2014). The RM observations have found
a relation between R$_{in}$ and UV luminosity as $R_{in} = \Delta t *c \propto
L_{UV}^{0.5}$ (Minezaki et al. 2004; Suganuma et al. 2006), where $\Delta t$ is
the lag between optical and IR flux variations and $c$ is the speed of light.  
For Seyfert galaxies  with  typical UV luminosities of 10$^{42}$ $-$  
10$^{44}$ ergs s$^{-1}$,   
the inner radius at which the hot dust sublimates can be of the order of 
0.01 $-$ 0.1 pc or 10 $-$ 100 light days  (Suganuma et al. 2006). 
Also, 
near-infrared (NIR) 
interferometric observations have been 
able to measure the size of the dusty torus in a few nearby AGN
(Swain et al. 2003, Kishimoto et al. 2009, 2011a, Pott et al. 2010, Weigelt
et al. 2012)

In addition to NIR, mid-IR interferometric observations are also
available for about two dozen AGN (Tristram et al. 2007; 
Burtscher et al. 2009; Tristram et al. 2009; Honig et al. 2013).  It has been pointed out by 
Kishimoto et al. (2007) that the $R_{in}$ of the dusty torus found by 
RM is systematically smaller by a factor of 3 compared to 
sublimation radius (R$_{sub}$) predicted from the
dust sublimation temperature (T$_{sub}$) of about 1600 K (Barvainis 
1992). It has been thought that this discrepancy may be due 
to a combination of many factors such as the dust grain size, clumpy dusty 
torus, anisotropic illumination of the accretion disk 
(Kawaguchi $\&$ Mori 2010), winds/outflows from the accretion disk
(Konigl $\&$ Kartje 1994; Elitzur $\&$ Shlosman 2006; 
Czerny $\&$ Hryniewicz 2011), etc. The observations 
to confront  predictions from T$_{sub}$ are available for 17 AGN monitored for dust reverberation. However the RM data for these 17 AGN have poor time resolution. As the sample size is small to obtain constrains on the nature of the torus as well as to understand the causes for the discrepancy between observations and 
predictions, it is very important to have RM estimates of torus size for many AGN. Towards this objective, we have started a monitoring project called the 
REverberation Mapping of Active Galactic Nuclei Program (REMAP). REMAP was 
started in the year 2016, and it involves carrying out optical V and B-band, and infrared J, H and $K_s$ band observations of a sample of AGN. As of now REMAP uses the 2m Himalayan Chandra Telescope at Hanle.

\section{The technique of Reverberation mapping}
Echo mapping or  Reverberation mapping is a standard tool for probing the structure and kinematics of the BLR and dust region in AGN. The emission line from BLR or the NIR flux from torus echoes with the variation in UV/optical fluxes from accretion disk. In its simplest form, the mean time delay between continuum and emission line/NIR variations is measured typically by cross-correlation of the respective light curves. Suppose a(t) denotes the UV/optical light curve of the accretion disk and b(t) represents the line emission light curve from BLR or NIR light curve from torus, then a(t) and b(t) are causally connected 
with a time lag $\tau$ through the transfer equation (Peterson 2001):
\begin{equation}
b(t) = \int \psi(\tau) a(t-\tau) d\tau
\end{equation} 
where $\psi(\tau)$ is the transfer function which encodes the geometry and 
kinematics of the emitting region. The RM technique relies on the following three important assumptions (Peterson 2001)
\begin{enumerate}
\item The continuum originates in a single central source of isotropic radiation.
\item Light-travel time is the most important time scale here.
\item {There is a simple, though not necessarily linear relationship between the observed continuum and the ionizing continuum.}
\end{enumerate}
Thus, by measuring a(t) and b(t), one can infer the geometry and kinematics
of the emitting region.
  
\section{The REMAP project}
Determination of the extent of the inner edge of the dusty torus in an AGN
through RM is not very expensive compared to interferometric
observations. Also, through reverberation mapping observations we can
determine the torus size in distant AGN, while interferometric 
observations are limited to just a few bright and nearby AGN. However, 
RM observations are challenging and the difficulty lies in accumulating 
sensitive and high cadence light curves. The method of RM as of now has been applied to about twenty AGN to measure 
the inner rim of the torus (Koshida et al. 2014). The possible reasons for the paucity of such data could be due to a) difficulties in scheduling observations 
in the optical and NIR, b) due to limited attempts made in this direction. Our 
objective through the REMAP project is to increase the number of AGN in the low-luminosity end of AGN dust torus size $-$ luminosity relation with the
RM based estimates of the dust torus size inferred from
high cadence optical IR monitoring observations. The initial sample
of sources for monitoring under REMAP were selected as follows
\begin{enumerate}
\item Firstly, we took the sample of 60 sources for  which RM of BLR were 
available (Bentz et al. 2015).  This is because, there are only a few sources for which the time lag 
measurements for both BLR and the torus are available. Comparisons of these 
measurements indicate that the torus lag is 4$-$5 times larger than the BLR lag.
\item Secondly, we imposed the condition that the BLR lag should be greater
than 3 days and lesser than 10 days. This condition was guided by our ability
to observe each source once every five days, which sets the time 
resolution of our data. With this time resolution it would be possible to 
obtain meaningful lag only for those sources that are expected to have dust
lag more than 30 days or so and monitored for a duration of about 6 months.
\item Thirdly the sources must be brighter than 16.5 magnitude in the V-band
and must have declination greater than $-$10 degree. The magnitude constraint
was imposed so as to get good S/N data within our allocated time at the HCT 
and the declination limit is for the observability of sources from HCT.
\end{enumerate}
The above selection criteria lead us to a final sample of 8 sources for 
REMAP. The details of these sources are given in Table 1.
 
\section{Observations, reduction and analysis}

\begin{table}
\caption{Details of the initial list of objects selected for the REMAP 
monitoring.
\label{table1}} 
\begin{center}
\resizebox{1.0\textwidth}{!}{ 
\begin{tabular}{l l r r l c c r r}
\hline 
No. & Object Name & RA (2000) & DEC (2000) &   V (mag) & $z$ & Type & M$_{BH}$ ($\times10^6M_{\odot}$) & R$_{BLR}$ (days) \\
\hline
1 & H0507+164     & 05:10:45.5 &   +16:29:56 & 15.64    & 0.018 &  Seyfert 1.5 & $9.62^{+0.33}_{-3.73}$ & $3.01^{+0.42}_{-1.84}$\\
2 & Z 229-15      & 19:05:25.9 &   +42:27:40 & 15.40    & 0.028  &  Seyfert 1 & $10.00^{+1.90}_{-2.40}$ & $3.86^{+0.69}_{-0.90}$  \\
3 & Mrk 142       & 10:25:31.3 &   +51:40:35 & 16.15    & 0.045 & Seyfert 1 & $2.17^{+0.77}_{-0.83}$ & $2.74^{+0.73}_{-0.83}$ \\
4 & MCG+10-16-111 & 11:18:57.7 &   +58:03:24 & 16.71   & 0.028 & Seyfert 1 & $5.50^{+2.00}_{-1.80}$ & $2.31^{+0.62}_{-0.49}$ \\
5 & Arp 151       & 11:25:36.2 &   +54:22:57 & 16.49    & 0.021 & Seyfert 1 & $6.41^{+0.92}_{-1.19}$ & $3.99^{+0.49}_{-0.68}$ \\
6 & Mrk 1310      & 12:01:14.3 & $-$03:40:41 & 15.91    & 0.019 & Seyfert 1 & $2.20^{+0.90}_{-0.90}$ & $4.20^{+0.90}_{-0.10}$ \\
7 & Mrk 202       & 12:17:55.0 &   +58:39:35 & 16.41    & 0.021 & Seyfert 1 & $1.30^{+0.40}_{-0.40}$ & $3.50^{+0.10}_{-0.10}$ \\
8 & NGC 5273      & 13:42:08.3 &   +35:39:15 & 13.12    & 0.003 & Seyfert 1 & $4.70^{+1.60}_{-1.60}$ & $2.21^{+1.19}_{-1.60}$\\
\hline 
\end{tabular} 
}
\end{center} 
\end{table}

\subsection{Observations}
REMAP monitoring started in the year 2016 and since then data are being
accumulated once every 5 days. Given the observational constraints, the 
strategy that is currently adopted is that one source  will be monitored
for a minimum duration of about six months. Observations are being carried out 
using 2-m Himalayan Chandra Telescope (HCT) operated by Indian Institute of Astrophysics, India. The optical observations are being carried out using
the Himalayan Faint Object Spectrograph and Camera (HFOSC) mounted at the 
Cassegrain focus equipped with a 2K$\times$4K SiTe CCD system. Each pixel of the 
CCD corresponds to 0.3 $\times$ 0.3 arcsec$^2$, covering a total field 
of view of  
10 $\times$ 10 arcmin$^2$. The typical exposure times in B and 
V-band are about 150$-$200 seconds and 50$-$100 seconds, respectively depending
on the sky conditions on any particular night of observation. The IR 
observations in J, H and $K_s$ bands are being made using the TIFR Near 
Infrared 
Spectrometer (TIRSPEC) mounted on one of the side ports of HCT 
(Ninan et al. 2014), covering a field of view of 5$\times$5 arcmin$^2$. The IR 
observations are performed in dithered mode at three dither positions each 
having 20 seconds exposure time for each of three IR filters namely J, H and 
$K_s$. Thus at any given epoch, observations in two optical bands
(B and V) and three IR bands (J, H and K$_s$) are completed within 45 minutes.

\subsection{Data reduction}

Reduction of the data acquired in optical B and V bands were carried out using
IRAF (Image Reduction and Analysis Facility)\footnote{IRAF is operated by the Association of Universities for Research in Astronomy, Inc., under cooperative agreement with the National Science Foundation.} and MIDAS (Munich Data Analysis 
System)\footnote{MIDAS is the trade-mark of the European Southern Observatory.}. As some of the source in our sample  are at low redshift, host galaxy is conspicuously present in the acquired image frames (as seen in Mrk 202), that can 
contaminate the photometry of the AGN. In such cases GALFIT (Peng et al. 2002) 
is used to remove the host galaxy contribution. The NIR data are reduced 
using TIRSPEC NIR Data Reduction Pipeline developed by Ninan et al. (2014). We used differential photometry to convert the instrumental magnitudes to apparent magnitudes using few stars present in the same image frame, whose apparent magnitudes are taken from SIMBAD{\footnote{http://simbad.u-strasbg.fr/simbad/}}. After correcting for the galactic extinction taken from the NASA/IPAC Extragalactic data base (NED){\footnote{https://ned.ipac.caltech.edu/}}, the apparent magnitudes were converted into fluxes.  

\subsection{Analysis}
\subsubsection{Subtraction of the accretion disk component from the NIR flux}
For sources that are monitored as part of REMAP, multi-band light curves are 
first generated. They are then cross-correlated to find the delay between
the optical V-band and three NIR bands. For this the NIR flux is expected to be
only from the torus. However, this is not the case as the NIR-flux also 
contains emission component from the accretion disk. The presence of 
accretion disk component of NIR-flux to the generated IR light curves 
would make the time lag shorter than the actual lag of the dust-torus emission 
(Koshida et al. 2014). Therefore, the contribution of accretion disk to 
observed NIR-flux values needs to be removed before any
correlation analysis between the light curves in V-band and NIR. For this
the contribution of accretion disk to the NIR flux is evaluated as 
follows (Koshida et al. 2014)

\begin{equation}
f_{NIR,disk}(t) = f_V(t) \big(\frac{\nu_{NIR}}{\nu_{V}}\big)^{\alpha_{\nu}}
\end{equation}   
where $f_{NIR,disk}(t)$ and $f_V(t)$ are accretion disk component of the NIR flux and the V-band flux at time t respectively. $\nu_V$ and $\nu_{NIR}$ are effective frequencies of V, NIR (J, H, $K_s$) bands respectively and $\alpha_{\nu}$ is the power-law index which can be estimated using 
\begin{equation}
\alpha_{\nu} = \frac{ln\big(f_B/f_V\big)}{ln\big(\nu_B/\nu_V\big)}
\end{equation}
The power law index is evaluated for each epoch for our observations using
the V and B-band measurements which is then used to correct for the contribution
of the accretion disk to the NIR flux.
\subsubsection{Lag between optical and NIR flux variations}
To have a quantitative estimate of the lag between optical and NIR flux 
variations, we use both the discrete cross-correlation function 
(Edelson $\&$ Krolik 1988) and the 
interpolated cross correlation function (Gaskell $\&$ Sparke 1986; 
Gaskell $\&$ Peterson 1987). Uncertainties in the derived lags are estimated
using Monte Carlo simulation methods that used both flux randomization
and random subset selection using the procedures outlined in 
Peterson et al. (1998).
\section{Results}
Since the start of REMAP, observations are regularly taken from the 2m HCT. We
briefly discuss the results obtained from this programme.
\subsection{Mrk 202}
Mrk 202 is a Seyfert 1 galaxy at a redshift of $z$ = 0.021. Monitoring of this
object started in the year 2016, however, as the observing run was affected by 
bad weather, we could accumulate 14 data points for 
Mrk 202 in optical V-band and NIR J-band. This sources has a prominent host 
galaxy in the observed V-band images (Figure 1). To get the flux of the AGN, 
GALFIT (Peng et al. 2002) was used to remove the contribution of the host 
galaxy for each epoch so that changes in seeing will not affect in the final flux-measurement. Aperture photometry
was done on the images after removal of the underlying host galaxy
component (Figure 2). The generated V and J-band light curves are shown in
Figure 3. To correct for the contribution of the accretion disk to the
observed J-band flux Eq. 2 was used wherein $\alpha$ was taken as 1/3 (Koshida et al. 2014, Mandal et al. 2018).
Due to the availability of limited amount of observations, time lag 
could not be obtained for Mrk 202. This object is planned for re-observation
from the middle of 2019.

\subsection{Z229-15}
Z229-15 is a Seyfert 1 galaxy at a redshift of $z$ = 0.028. About 25 epochs
of observations were accumulated in the optical V and B-bands and in the
NIR J, H and K$_s$ bands. The contribution of accretion disk to the NIR flux
was corrected as per the details outlined in Section 4.3.1. Preliminary 
light curves in the optical B-band and NIR K$_s$ band are shown in Figure 4.
Further analysis of the data acquired on this object is in progress.

\subsection{H0507+164}
About 35 epochs of data were accumulated for this object in the optical V-band
and NIR J,H and K$_s$ bands. The accumulated data were analysed
following the procedures given in Section 4.2 and 4.3. Using cross-correlation
analysis we found a lag of 34.6$^{+12.1}_{-9.6}$ days between V and
K$_s$ bands. Thus from light travel arguments, we found the inner radius
of the dust torus in H0507+164 as 0.03$^{+0.01}_{-0.01}$ pc 
(Mandal et al. 2018).

\begin{figure}[h]
\begin{minipage}{8cm}
\centering
\includegraphics[width=5cm]{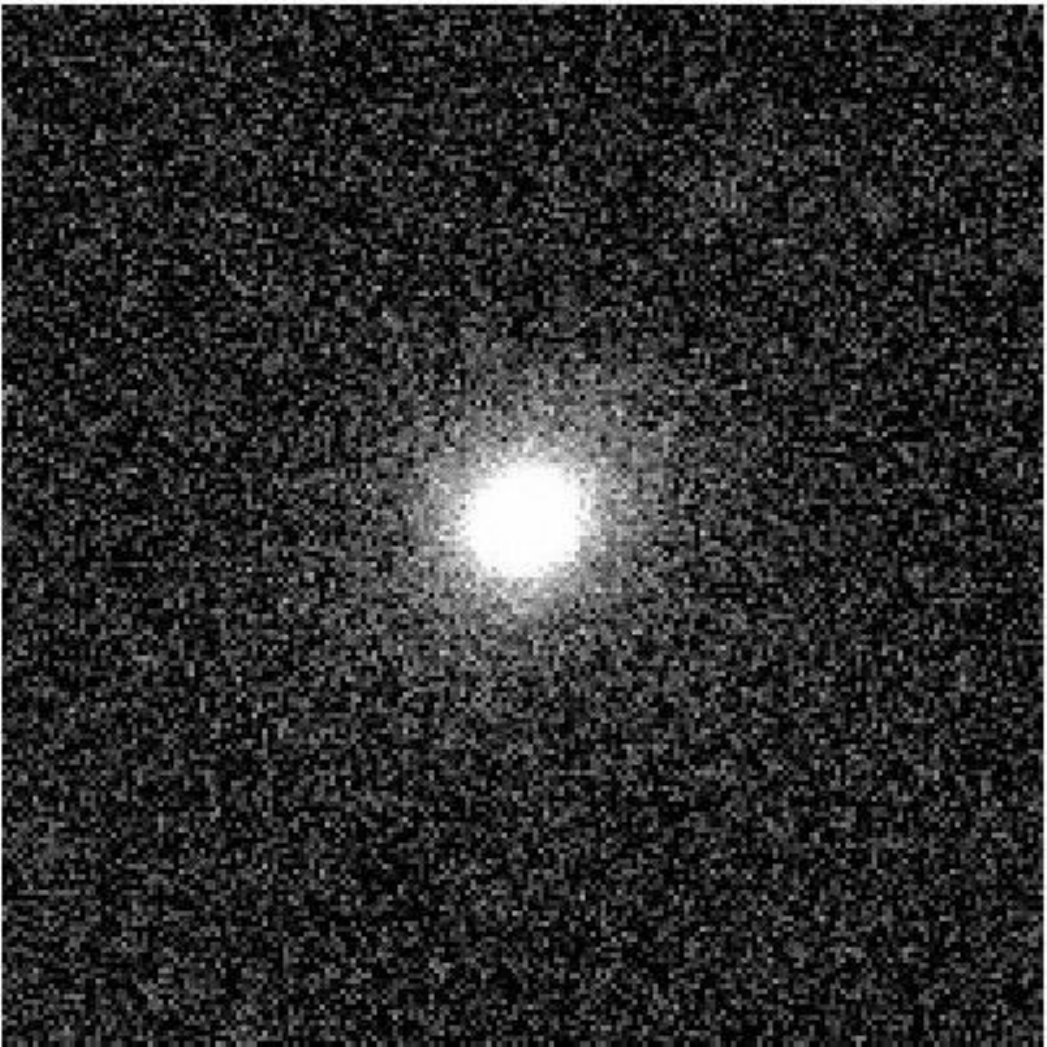}
\caption{Bias-subtracted $\&$ flat-fielded observed image of Mrk 202 in V-band.\label{fig_1}}
\end{minipage}
\hfill
\begin{minipage}{8cm}
\centering
\includegraphics[width=5cm]{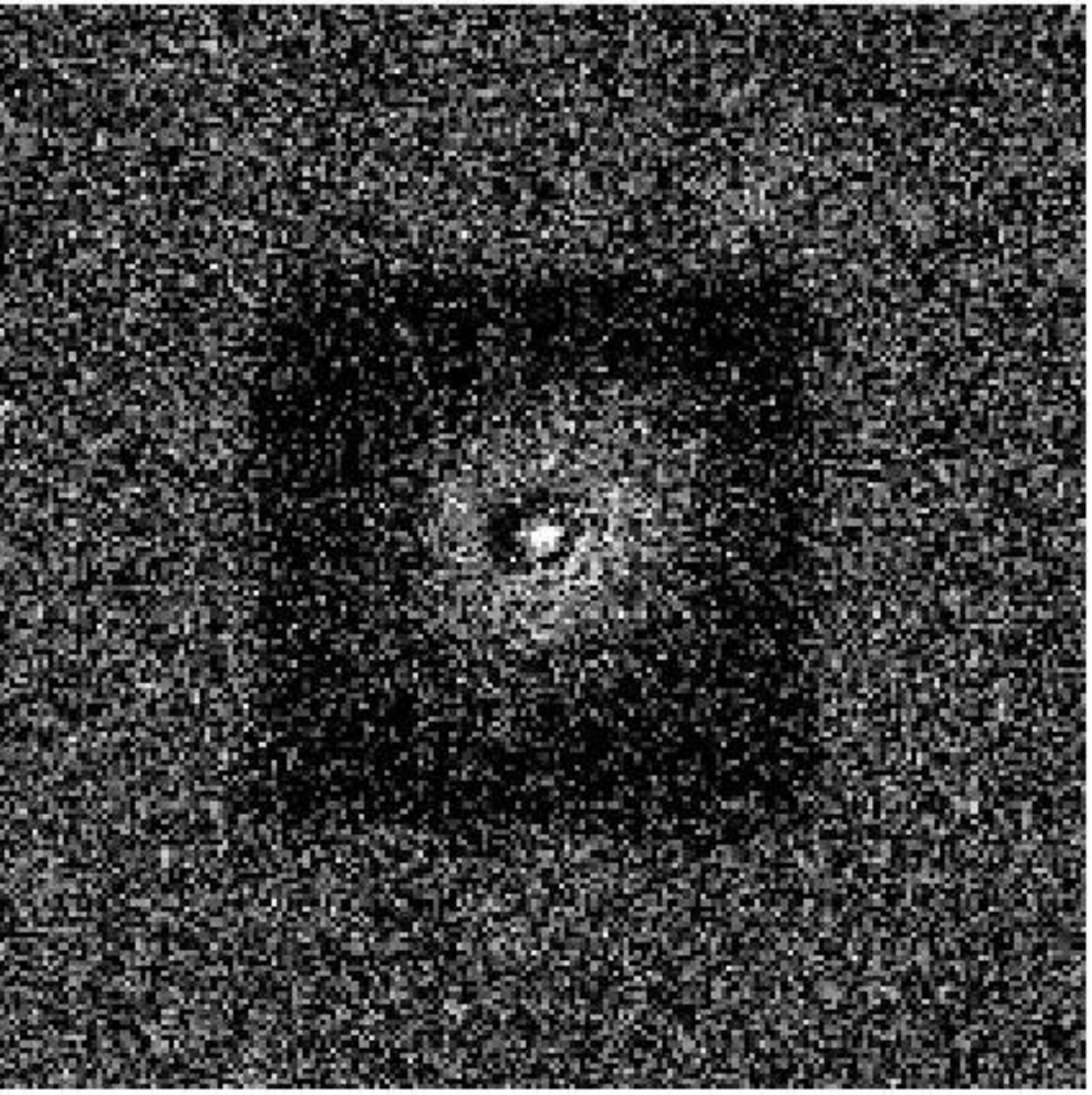}
\caption{Residual image of Mrk 202 obtained using GALFIT. \label{fig_2}}
\hfill
\end{minipage}
\end{figure}

\begin{figure}[h]
\begin{minipage}{8cm}
\centering

\includegraphics[width=7.8cm]{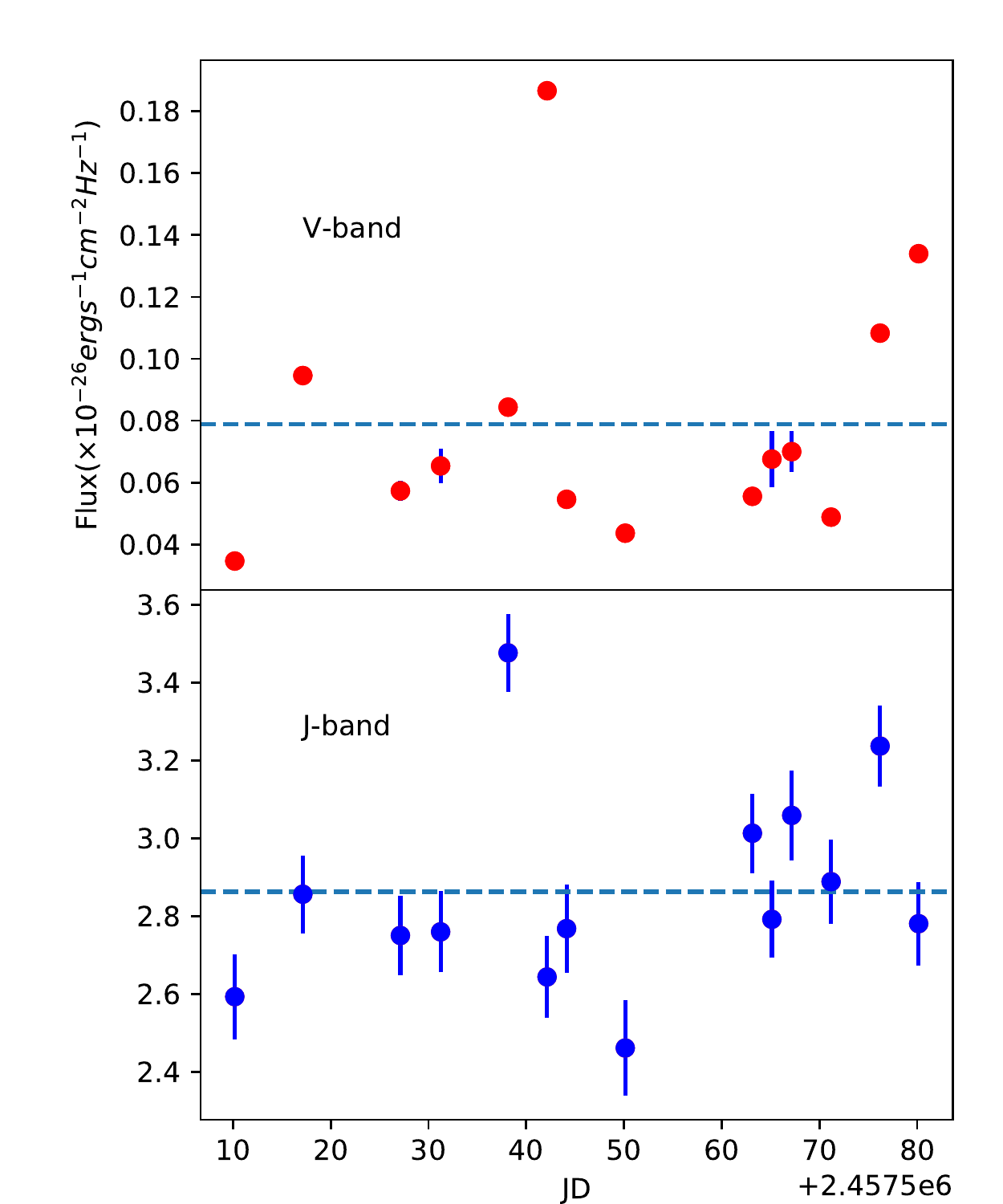}
\caption{Light curves for Mrk 202. The J-band fluxes are corrected for accretion disk contribution.\label{fig_3}}

\end{minipage}
\hfill
\begin{minipage}{10cm}
\centering
\includegraphics[width=8cm]{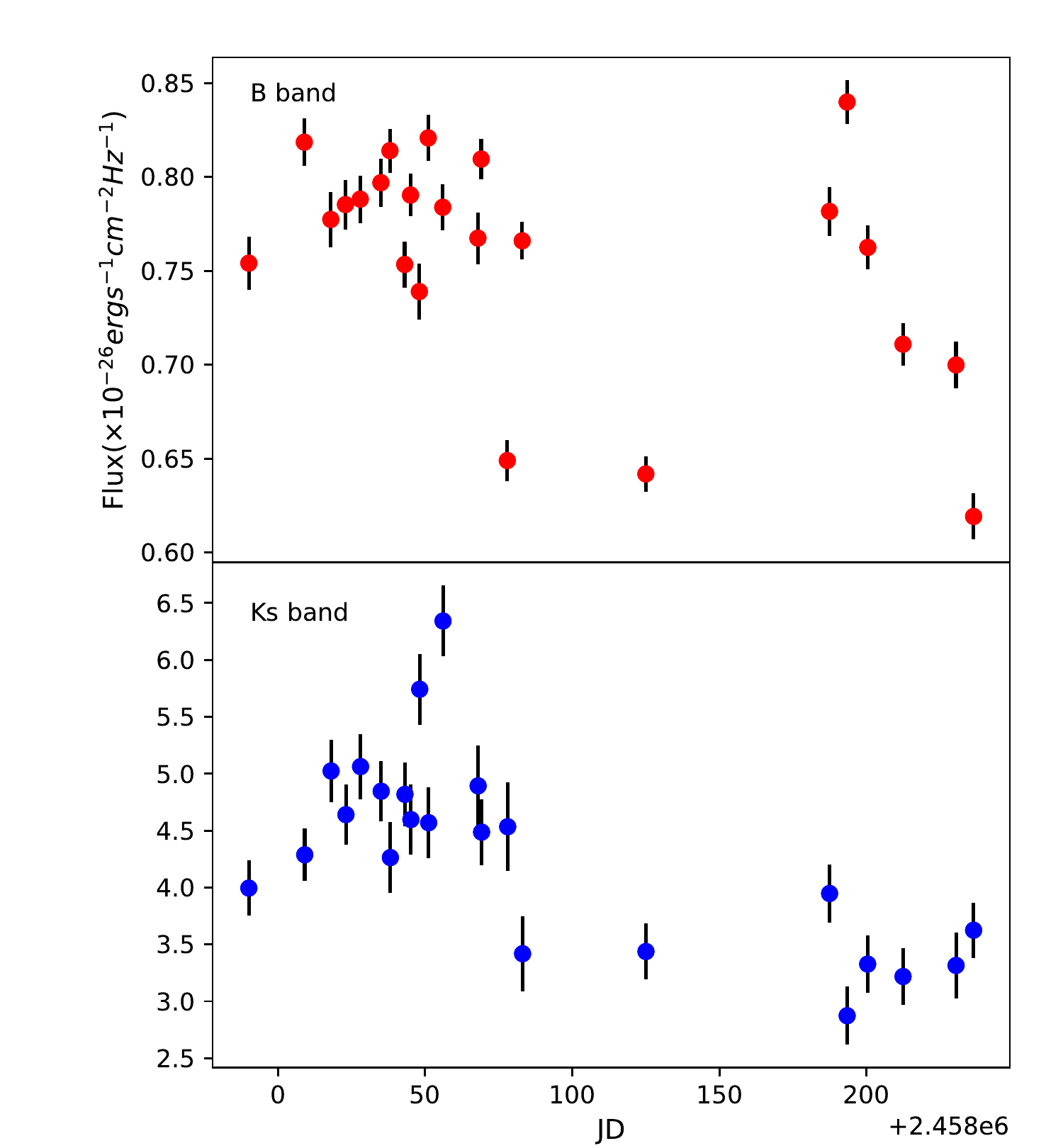}
\caption{The light curves of Z229-15 in B and $K_s$-bands. The fluxes are not corrected for host-galaxy contribution. The $K_s$-band fluxes were corrected for the contamination of emission from the accretion disk.\label{fig_4}}
\end{minipage}
\end{figure}

\section{Summary}
REMAP was started as an observational program on the 2m HCT, with the prime
motivation to find the inner edge of the dusty torus in more number of AGN
using the technique of RM. Towards this, our carefully selected
sample consists of a total of eight sources. We summarize the results obtained from REMAP till
today
\begin{enumerate}
\item We have successfully obtained the size of the torus in one source,
namely, H0507+164 a Seyfert 1.5 galaxy at $z$ = 0.018. Using about 35 epochs 
of data obtained between the period October 2016 to April 2017, we found that 
the inner edge of the torus is located at a distance of 0.029 pc from the 
central optical continuum source.
\item For Mrk 202, we could accumulate only 14 epochs of data. Though the 
observations has indicated the source is variable, the acquired data
are not sufficient enough to determine the time lag via cross-correlation
analysis. This object is planned for observation in the upcoming cycle in 
2019.
\item For Z229-15, observations are over. Analysis of the data indicates
that the sources is variable. Further analysis is in progress. By the end of
the year 2021, we expect to have the radius of the inner edge of the torus
in all the eight AGN selected for REMAP. Efforts are also being taken to 
increase the sample size and involve other medium size telescopes.
\end{enumerate}

%
%
\section*{Acknowledgements}
We thank to the supporting staff at the Indian Astronomical Observatory (IAO), Hanle, and CREST, Hoskote. AKM and RS thank the National Academy of Sciences, India for providing the required fund for this project. This work also used the SIMBAD data base, operated at CDS, Strasbourg, France, the NED, which is operated by the Jet Propulsion Laboratory, California Institute of Technology, under contract with NASA and data products from the Two Micron All Sky Survey, a joint project of the University of Massachusetts and the Infrared Processing and Analysis Center/California Institute of Technology, funded by the NASA and the National Science Foundation. AKM acknowledges the local support from the organising committee of the 2nd Belgo-Indian Network for Astronomy $\&$ astrophysics (BINA) workshop, held in Brussels, Belgium in October, 2018.
%
%
%

\footnotesize
\beginrefer
\refer Antonucci R. 1993, ARA$\&$A, 31, 473

\refer Barvainis R. 1992, ApJ, 400, 502

\refer Bentz M. C., Katz S. 2015, PASP, 127, 67

\refer Blandford R. D., McKee C. F. 1982, ApJ, 255, 419

\refer Burtscher L., Meisenheimer K., Tristram K. R. W. et al.  2013, A$\&$A, 558, 149

\refer Clavel J.,  Wamsteker W., Glass I. 1989, ApJ, 337, 236

\refer Czerny B., Hryniewicz K. 2011, A$\&$A, 525, L8

\refer Edelson R. A., Krolik J. H. 1988, ApJ, 333, 646

\refer Elitzur M., Shlosman I. 2006, ApJ, 648, L101

\refer Gaskell C. M., Peterson B. M. 1987, ApJS, 65, 1

\refer Gaskell C. M., Sparke L. S. 1986, ApJ, 305, 175

\refer Glass I. S. 2004, MNRAS, 350, 1049

\refer Honig S. F., Kishimoto M., Tristram K. R. et al. 2013, ApJ, 771, 87

\refer Kawaguchi T., Mori M. 2010, ApJ, 724, L183

\refer Kishimoto M., Honig S. F., Antonucci R. et al. 2009, A$\&$ 507, L57

\refer Kishimoto M., Honig S. F., Antonucci R. et al. 2011a, A$\&$A, 527, A121

\refer Kishimoto M., Honig S. F., Antonucci R. et al. 2011b, A$\&$A, 536, A78

\refer Konigl A., Kartje J. F. 1994, ApJ, 434, 446

\refer Koshida S., Minezaki T., Yoshii Y.  et al. 2014, ApJ, 788, 159

\refer Koshida S., Yoshii Y., Kobayashi Y. et al. 2009, ApJL, 700, L109

\refer Mandal A. K., Rakshit S., Kurian K. S. et al. 2018, MNRAS, 475, 5330

\refer Minezaki T., Yoshii T., Kobayashi Y. et al. 2004, ApJL, 600, L35

\refer Nelson B. O. 1996, ApJL, 465, L87

\refer Ninan J. P., Ojha D. K., Ghosh S. K. et al. 2014, JAI, 3, 1450006

\refer Peng C. Y.,  Ho L. C., Impey C. D. et al. 2002, AJ, 124, 266

\refer Peterson B. M., Wanders I., Horne K. et al. 1998, PASP, 110, 660

\refer Peterson B. M. 2001, Advanced Lectures on the Starburst-AGN Connection,  3 

\refer Pott J. U., Malkan M. A., Elitzur M.  et al. 2010, ApJ, 715, 736

\refer Suganuma M., Yoshii Y., Kobayashi Y. et al. 2006, ApJ, 639, 46

\refer Swain M., Vasisht G., Akeson R. et al. 2003, ApJL, 596, 163

\refer Tristram K. R. W., Meisenheimer K., Jaffe W. et al. 2007, A$\&$A, 474, 837

\refer Tristram K. R. W., Raban D., Meisenheimer K. et al. 2009, A$\&$A, 502, 67

\refer Urry C. M., Padovani P. 1995, PASP, 107, 803

\refer Weigelt G., Hofmann K. H., Kishimoto M. et al. 2012, A$\&$A, 541, A9

\endrefer

\end{document}